\documentclass{article}

\usepackage{arxiv}

\usepackage[utf8]{inputenc} 
\usepackage{hyperref}       
\usepackage{url}            
\usepackage{booktabs}       
\usepackage{amsmath}
\usepackage{amsfonts}       
\usepackage{amssymb}
\usepackage{microtype}      
\usepackage{graphicx}
\usepackage{natbib}
\usepackage{doi}
\usepackage{authblk}
\usepackage{algorithm}
\usepackage{algorithmic}
\usepackage{float}
\usepackage{subcaption}
\usepackage{tikz}
\usetikzlibrary{arrows.meta, positioning}

\title{Development of an Effective Federated Learning-Based Defense System Against Phishing Attacks on Mobile Platforms}

\date{}

\author[1]{Eren Dogan}
\author[1]{Busra Buyuktanir}
\author[2]{Gozde Karatas Baydogmus\thanks{Corresponding author: \texttt{gkaratasbaydogmus@luc.edu}}}
\author[1]{Kazim Yildiz}
\affil[1]{Department of Computer Engineering, Faculty of Technology, Marmara University, Istanbul, Turkiye}
\affil[2]{Department of Computer Science, Loyola University Chicago, Chicago, USA}

\renewcommand{\shorttitle}{Federated CNN Defense Against Mobile Phishing}
\renewcommand{\headeright}{}
\renewcommand{\undertitle}{}

\hypersetup{
pdftitle={Development of an Effective Federated Learning-Based Defense System Against Phishing Attacks on Mobile Platforms},
pdfauthor={Eren Dogan, Busra Buyuktanir, Gozde Karatas Baydogmus, Kazim Yildiz},
pdfkeywords={Mobile security, Federated learning, Phishing detection, Privacy-preserving AI, Real-time protection},
}

\begin{document}
\maketitle

\begin{abstract}
Phishing remains one of the most pervasive and damaging cyber threats, exploiting both technical flaws and human behavior to compromise sensitive data. We present a privacy-preserving, real-time defense that analyzes full webpage source code on mobile platforms. Our system combines a lightweight Convolutional Neural Network (CNN) for end-to-end feature extraction with a Federated Learning (FL) framework that updates models collaboratively on user devices without transferring raw data-supporting regulatory compliance and personalization. Evaluated on a large and diverse dataset, the model attains 86.02\% accuracy, 90.35\% precision, 80.64\% recall, and 92.78\% ROC-AUC. While some centralized, classical approaches may report higher accuracy, our design emphasizes a practical balance among accuracy, user privacy, and real-time performance on resource-constrained devices. We further demonstrate an end-to-end integration within the open-source Lightning Android browser, achieving real-time detection with low computational overhead. This work shows that privacy-first, AI-driven phishing detection is both feasible and scalable across mobile and IoT settings, delivering robust protection while keeping user data on-device.
\end{abstract}


\section{Introduction}

The widespread adoption of information technologies has significantly transformed communication and data exchange, but it has also expanded the cyber attack surface. In Turkiye, household internet access reached 96.2\% and internet usage among individuals aged 16--74 rose to 90.9\% in 2025 \citep{tuik2025}, contributing to rising incidents of data breaches, malware, and particularly phishing attacks \citep{omer2023,bicakci2016}.

Phishing remains a major threat by exploiting both technical vulnerabilities and human behavior. Reports from the Anti-Phishing Working Group (APWG) show that global phishing volumes have continued to climb, exceeding 1.1 million attacks in a single quarter of 2025---among the highest quarterly totals ever recorded \citep{apwg2024, apwg2025}. These attacks typically deceive users into revealing sensitive information (e.g., credentials or financial data) via fraudulent emails, messages, or websites \citep{unver2011,hekim2015,buber2017}. Recent studies show that 85\% of organizations experienced phishing incidents within a year, often leading to financial and reputational damage \citep{ukcsbs2025,wassermann2023,sharif2022}.

While phishing detection has been extensively studied for desktop environments, mobile platforms introduce additional challenges. Limited screen size, simplified interfaces, device heterogeneity, and constrained resources reduce the effectiveness of traditional detection methods \citep{wu2015, canova2015, varshney2016}. Existing countermeasures include user education, blacklist-based filtering, and machine learning approaches \citep{aleroud2017, jain2022}. However, blacklists quickly become outdated, and centralized ML models raise privacy concerns due to the need for aggregating user data.

Federated Learning (FL) offers a promising alternative by enabling collaborative model training without sharing raw data, thereby preserving user privacy while improving generalization \citep{yang2019federated,buyuktanir2025federated,buyuktanir2024ubak,buyuktanir2025ddos_fl}. When combined with deep learning models such as Convolutional Neural Networks (CNNs), which can extract hierarchical patterns directly from web page source code, FL enables robust and privacy-aware phishing detection.

Despite its advantages, deploying FL in mobile environments introduces challenges, including limited computational resources, communication overhead, device heterogeneity, and security risks such as model poisoning. Addressing these constraints requires lightweight architectures and efficient training strategies.

In this study, we propose a real-time, CNN-based phishing detection system integrated with federated learning for mobile platforms. The system performs automatic feature extraction from HTML source code while ensuring that sensitive data remain on-device. Although designed for mobile use, the proposed framework is also applicable to IoT environments, where privacy-preserving, distributed intelligence is increasingly critical \citep{nguyen2021federated,ozturk2025differential}.

\subsection{Objectives and Contributions}

This paper proposes a privacy-preserving, real-time phishing detection framework leveraging federated learning and deep neural networks. Unlike many previous works that either focus on pure centralized detection or rely on limited feature resources, this study emphasizes on-device learning, adaptive feature extraction from full source code, and scalability in architecture to IoT infrastructures. The main objectives and contributions of the work are summarized below.

\begin{itemize}
    \item To design a federated learning framework that allows continuous adaptive model training across mobile devices without violating user privacy.
    \item To propose a lightweight CNN model capable of extracting deep, dynamic features from complete web page source code, ensuring robustness against obfuscation and zero-day phishing patterns.
    \item To validate the proposed framework on real-world, large-scale datasets and compare it against both centralized and classical ML baselines concerning practical privacy and resource constraints for mobile deployment.
    \item To analyze federated performance trade-offs, including communication cost, model robustness, and privacy--utility balance, and provide practical insights for deployment in mobile and IoT environments.
\end{itemize}

\subsection{Research Questions and Hypotheses}

This study addresses the following research questions:
\begin{enumerate}
\item How effectively can a lightweight CNN detect phishing from web page source code on mobile devices (accuracy, precision, recall)?
\item Can the model be deployed in a mobile browser for real-time detection under constraints of latency and model size?
\item Does federated learning improve generalization and detection performance compared to centralized training?
\item How does the proposed CNN+FL approach overcome limitations of shallow-feature and centralized mobile phishing detection systems?
\end{enumerate}

Based on these questions, we hypothesize that: (i) a lightweight CNN combined with efficient FL can meet mobile resource constraints; (ii) the federated model achieves competitive performance (\(\geq 85\%\) accuracy) while preserving privacy; (iii) FL enables personalization and adaptability to user behavior; and (iv) secure aggregation and client-side optimization mitigate poisoning risks and communication overhead.

The remainder of the paper is organized as follows: Section~\ref{sec:related} reviews related work; Section~\ref{sec:method} presents the methodology; Section~\ref{sec:results} reports experimental results; and Section~\ref{sec:conclusion} concludes the study.

\section{Related Work}\label{sec:related}
Phishing attack detection is one of the most actively explored topics in cybersecurity literature, owing to the growing diversity of attack strategies and the rapid expansion of mobile device usage \citep{sun2022}. In this section, we present an overview of the main approaches and advancements in the field, focusing on (i) traditional and machine learning-based methods, (ii) deep learning developments, (iii) phishing detection on mobile platforms, (iv) the role of federated learning and data privacy, and (v) the specific research gap that this study aims to address.

\subsection{Classical and Machine Learning-Based Phishing Detection}

Early automated phishing detection systems primarily relied on blacklists, whitelists, and rule-based approaches \citep{zhang2007}. Although these techniques were effective against known threats, they quickly became outdated as attackers developed new, unreported methods, leading to a gradual decline in detection rates over time \citep{sun2022}. Consequently, researchers began adopting machine learning techniques that leverage multidimensional features, including statistical properties of URLs, domain structures, the presence of HTTPS certificates, page forms, hyperlinks, and both HTML and JavaScript content \citep{mohammad2014, jain2022}.

Large-scale open datasets such as the UCI Machine Learning Repository, PhishTank, OpenPhish, and Alexa Top Sites have facilitated the training and benchmarking of models like Decision Trees, Random Forests, and Support Vector Machines (SVMs), achieving detection accuracies exceeding 95\% in numerous studies \citep{dou2017, jain2023}. The integration of HTML and Document Object Model (DOM)-based features alongside static URL and text features has further enhanced performance. For instance, the CANTINA algorithm applies Term Frequency--Inverse Document Frequency (TF--IDF) analysis to assess HTML keyword density and has reported content-based detection rates as high as 94--97\% \citep{zhang2007}. Subsequent studies expanded these feature sets by incorporating address bar indicators, page content, form structures, and script analysis, thereby improving the robustness of detection systems against evolving threats \citep{jain2022}.

\subsection{Advances in Deep Learning for Phishing Detection}

In recent years, deep learning models have become a new benchmark in phishing detection. Architectures such as Convolutional Neural Networks (CNNs), Recurrent Neural Networks (RNNs), Long Short-Term Memory (LSTM), and Graph Convolutional Networks (GCNs) can automatically learn complex structural and semantic relationships, offering superior accuracy and adaptability compared to traditional machine learning techniques \citep{jain2023, islam2020, rao2023}.

CNN-based models directly analyze HTML, DOM, JavaScript, and visual features, using n-gram representations and layout information to differentiate phishing websites from legitimate ones \citep{islam2020, rao2023}. Notably, models like HTMLPhish have achieved detection rates exceeding 97\% through HTML-only analysis \citep{opara2020htmlphish}. GCN-based methods, on the other hand, exploit the graph structure of a page's DOM tree, achieving up to 7\% higher accuracy than classical approaches \citep{jain2023}.

Another research direction focuses on visual similarity-based detection, with models such as PhishZoo and Goldphish evaluating how closely fake websites mimic the visual design of legitimate ones---offering an additional protective layer \citep{jain2022, dou2017}.

\subsection{Phishing Detection on Mobile Platforms}

Phishing attacks targeting mobile devices have increased dramatically in recent years \citep{ndibwile2023}. Factors such as small screen sizes, reduced user attention, and hidden address bars provide attackers with new ways to deceive users. Most existing mobile-focused studies rely on lightweight machine learning algorithms, using features derived from URLs, basic HTML, and form elements to support efficient, real-time detection \citep{dhanavanthi2021, routhu2019}.

In both Turkish and international research, phishing detection on mobile devices using HTML/DOM features---with models such as Decision Tree, SVM, Random Forest, and rule-based filters---has achieved accuracy rates exceeding 90\% \citep{colhak2024, akpinar2022}. Innovative Android applications like UnPhishMe have introduced new techniques for identifying fake login forms and verifying page authenticity \citep{ndibwile2023}. However, while many existing mobile detection systems rely on features drawn from URLs, form elements, and basic HTML properties, fewer studies have explored automatically extracting complex and dynamic representations from full web page source code using end-to-end deep learning models. Applying such comprehensive, real-time analysis on mobile platforms---particularly within a privacy-preserving federated learning framework---remains a major gap in current research.

\subsection{Data Privacy with Federated Learning in Mobile and IoT Systems}

One of the most groundbreaking approaches in recent years is FL, which enables collaborative model training directly on user devices without uploading raw data to a central server \citep{buyuktanir2023cba, buyuktanir2026enhancing, cciplak2025fedetect, hard2018, li2023, thapa2023}. In FL, model updates are encrypted and aggregated, preserving user privacy and minimizing the risk of data exposure. Google's implementation of FL in its Gboard keyboard app has demonstrated substantial improvements in personalization and security while maintaining user privacy \citep{hard2018}.

For phishing detection, FL provides additional advantages, including real-time adaptability to new attack patterns and user behaviors, while maintaining the confidentiality of sensitive data \citep{li2023, ahmad2022}. Multiple studies have compared FL-based phishing detection with conventional centralized training, often finding that FL achieves similar or even superior accuracy with stronger privacy guarantees \citep{thapa2023}.

The decentralized and privacy-preserving characteristics of FL make it particularly suitable for Internet of Things (IoT) infrastructures, which consist of numerous distributed, resource-limited devices generating sensitive data across different environments \citep{zhang2022federated}. IoT systems frequently involve data such as health metrics, location information, household activity, and environmental sensor readings---making privacy protection essential \citep{perera2015big}. Traditional centralized approaches expose this data to greater privacy risks, whereas FL enables local model training on devices without transferring raw information, thereby minimizing exposure.

By employing FL, edge devices like smart home appliances, wearable technologies, industrial IoT components, and gateways can collaboratively train models while keeping user data on-device. This approach not only reduces communication overhead but also maintains strong privacy safeguards. Furthermore, FL removes the dependence on centralized servers, increasing resilience against common IoT-specific threats such as eavesdropping \citep{liao2018eavesdropping}, data manipulation \citep{covington2013threat}, and model poisoning attacks \citep{baracaldo2018}. The architecture supports compliance with regulations such as GDPR and KVKK while enhancing security and scalability across decentralized networks. Therefore, FL serves as a promising foundation not only for mobile-based systems but also for developing robust, privacy-aware AI solutions across a wide range of IoT ecosystems.

\subsection{Research Gap and Contribution of This Study}

Despite substantial progress, our literature review identifies a critical gap: the absence of a system that jointly provides (i) real-time, personalized phishing detection, (ii) deep and dynamic feature extraction from the complete page source via CNNs, (iii) robust operation on mobile platforms, and (iv) a privacy-preserving federated learning framework for continual adaptation. Prior work makes important advances, yet most studies either focus on desktop-based CNN analysis of page code or on mobile implementations that, while efficient, rely on hand-crafted, shallow features rather than end-to-end learning on the full source code \citep{jain2023, islam2020, colhak2024, akpinar2022}.

This study addresses that gap by introducing an integrated mobile solution that performs CNN-based dynamic feature extraction directly from page source code, supports privacy-preserving federated learning, and delivers real-time, personalized anti-phishing protection. By unifying these previously separate dimensions, our approach is practical and scalable, with strong potential for real-world deployment and societal impact.

\section{Methodology}\label{sec:method}

This study proposes a privacy-preserving phishing detection system that operates on mobile devices using CNN-based feature extraction and federated learning (FL). The system performs real-time analysis of web page source code while keeping both inference and training on-device.

\subsection{System Overview}

The system analyzes visited web pages in real time by processing their source code through a lightweight CNN model. Federated learning enables collaborative model updates without sharing raw data, ensuring privacy while adapting to evolving threats and user behavior.

\subsection{Research Design}

The workflow consists of the following stages:
\begin{enumerate}
\item \textbf{Data Preparation:} Collect, clean, and tokenize source code from phishing and legitimate websites.
\item \textbf{Model Development:} Design a CNN to extract discriminative features directly from source code.
\item \textbf{Optimization \& Training:} Tune hyperparameters via Bayesian optimization and evaluate using standard metrics.
\item \textbf{Federated Learning:} Enable on-device training and global model aggregation.
\item \textbf{Deployment:} Integrate the model into the \emph{Lightning} browser for real-time detection.
\item \textbf{Evaluation:} Measure latency, resource usage, and accuracy across devices.
\end{enumerate}

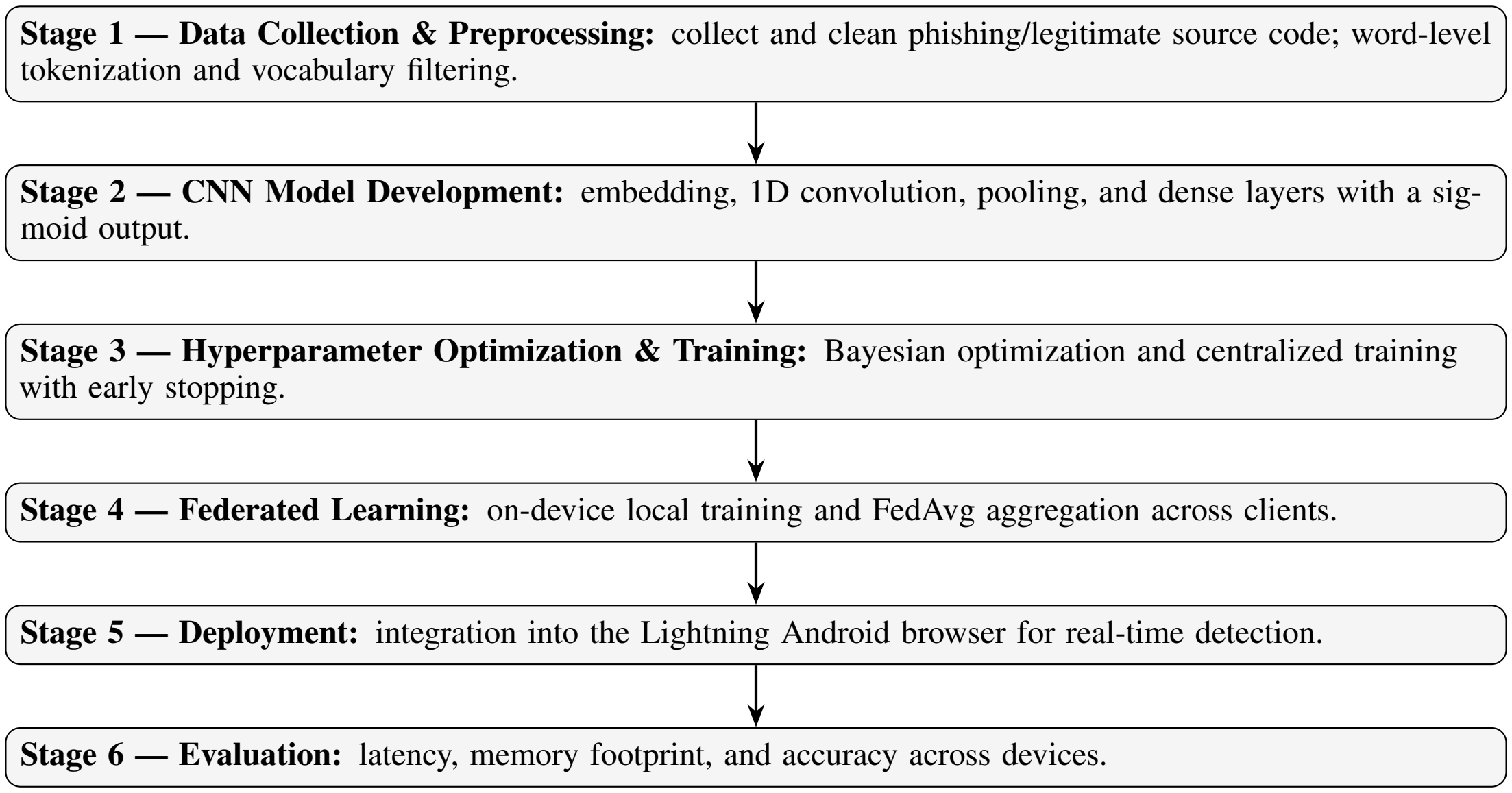
\begin{figure}[ht]
  \centering
  \begin{tikzpicture}[
    node distance=6mm and 0mm,
    stage/.style={rectangle, draw, rounded corners, align=left, text width=0.86\linewidth,
                  inner sep=4pt, font=\footnotesize, fill=gray!8},
    arr/.style={-{Stealth[length=2.2mm]}, thick}]
    \node[stage] (s1) {\textbf{Stage 1 --- Data Collection \& Preprocessing:} collect and clean phishing/legitimate source code; word-level tokenization and vocabulary filtering.};
    \node[stage, below=of s1] (s2) {\textbf{Stage 2 --- CNN Model Development:} embedding, 1D convolution, pooling, and dense layers with a sigmoid output.};
    \node[stage, below=of s2] (s3) {\textbf{Stage 3 --- Hyperparameter Optimization \& Training:} Bayesian optimization and centralized training with early stopping.};
    \node[stage, below=of s3] (s4) {\textbf{Stage 4 --- Federated Learning:} on-device local training and FedAvg aggregation across clients.};
    \node[stage, below=of s4] (s5) {\textbf{Stage 5 --- Deployment:} integration into the Lightning Android browser for real-time detection.};
    \node[stage, below=of s5] (s6) {\textbf{Stage 6 --- Evaluation:} latency, memory footprint, and accuracy across devices.};
    \draw[arr] (s1)--(s2); \draw[arr] (s2)--(s3); \draw[arr] (s3)--(s4);
    \draw[arr] (s4)--(s5); \draw[arr] (s5)--(s6);
  \end{tikzpicture}
  \caption{Overview of the proposed federated, CNN-based phishing detection pipeline. The workflow is organized into \emph{stages} (distinct from the federated-learning communication rounds used during training), spanning data preparation, model development and optimization, federated training, on-device deployment, and evaluation.}
  \label{fig:flowchart}
\end{figure}

An overview is shown in Figure~\ref{fig:flowchart}.

\subsection{Data Collection and Preprocessing}

\textbf{Data Sources:} Datasets were compiled from multiple repositories, including Alexa Top Sites, Google Search results, PhishTank, OpenPhish, PhishStats, Kaggle datasets, and a benchmark dataset \citep{vandooremaal2021dataset}.

\textbf{Preprocessing:} Because virtually every token in web source code can be informative, we deliberately omit the stop-word removal and stemming steps of traditional NLP pipelines so as to preserve the structural information of the code. Normalization consists of lowercasing, removing punctuation and non-alphanumeric symbols, and collapsing repeated whitespace; multiple character encodings (e.g., UTF-8, Latin-1) are handled during file reading.

\textbf{Tokenization:} We use a \emph{word-level} tokenizer. After normalization, the source code is split into word tokens and a frequency-ranked word index is built. A token is retained only if it is between 3 and 20 characters long and occurs at least 60 times in the corpus; overly rare, short, or long tokens are discarded. Each document is then mapped to a sequence of token indices and padded or truncated to a fixed length of 500 tokens. This reduces the vocabulary from roughly 20 million to about 406{,}000 tokens---filtering $\approx$98\% of the unique tokens---while retaining the most informative, structure-bearing terms.

\subsection{Model Architecture}

A lightweight CNN architecture is employed due to its balance between performance and efficiency. It consists of:
\begin{itemize}
\item embedding layer
\item 1D convolutional layers
\item max and global pooling
\item dense layers with dropout
\item sigmoid output for binary classification
\end{itemize}

The input consists of tokenized source code sequences, and the output is a binary label (phishing vs.\ legitimate).

\subsection{Hyperparameter Optimization}

Bayesian optimization is used to efficiently tune key hyperparameters (e.g., learning rate, filter size, network depth), improving performance while reducing computational cost.

The best-performing configuration identified by Bayesian optimization is summarized in Table~\ref{tab:cnn_hyperparams}.

\begin{table}[htp]
\centering
  \caption{CNN model hyperparameters.}
  \label{tab:cnn_hyperparams}
  \begin{tabular}{@{}ll@{}}
    \toprule
    Parameter & Value \\ \midrule
    Maximum sequence length & 500 \\
    Embedding vector size & 64 \\
    Number of Conv1D filters & 128 \\
    Kernel size & 5 \\
    Dense units (ELU activation) & 64 \\
    Dropout rate & 0.2 \\
    Learning rate & 0.0001 \\
    Mini batch size & 32 \\
    Epochs & Up to 50 (with early stopping) \\ \bottomrule
  \end{tabular}
\end{table}

This configuration provided a balanced, stable performance across both training data and real-world test sets.

\subsection{Model Training and Evaluation}

\subsubsection{Model Training}

The model was implemented and trained in TensorFlow within a Jupyter Notebook environment. To obtain reliable estimates and curb overfitting, the dataset was split into training, validation, and test sets. We used the Adam optimizer for its adaptive learning-rate updates and fast convergence. Early stopping served as regularization by monitoring validation loss and halting training when no improvement was observed for a predefined patience window. All training, hyperparameter searches, and evaluations were run on a high-performance Apple M2 Pro laptop.

\subsubsection{Model Evaluation}

We evaluated performance using complementary metrics that capture both accuracy and robustness, the classification performance metrics used in this study are summarized in Table~\ref{tab:metrics}.

\begin{table}
\centering
\caption{Classification performance metrics}
\label{tab:metrics}
\small
\begin{tabular}{p{3.2cm} p{9.5cm}}
\toprule
\textbf{Metric} & \textbf{Definition / Formula} \\
\midrule

\textbf{TP, TN, FP, FN} &
TP: correctly predicted positive \newline
TN: correctly predicted negative \newline
FP: false positive (Type I error) \newline
FN: false negative (Type II error) \\

\textbf{Accuracy} &
$\displaystyle \frac{TP + TN}{TP + TN + FP + FN}$ \\

\textbf{Precision} &
$\displaystyle \frac{TP}{TP + FP}$ \\

\textbf{Recall} &
$\displaystyle \frac{TP}{TP + FN}$ \\

\textbf{F1-score} &
$\displaystyle 2 \cdot \frac{\text{Precision} \cdot \text{Recall}}{\text{Precision} + \text{Recall}}$ \\

\textbf{ROC-AUC} &
$\displaystyle \int_{0}^{1} \mathrm{TPR}\, d(\mathrm{FPR})$ \newline
$TPR = \frac{TP}{TP + FN}, \quad FPR = \frac{FP}{FP + TN}$ \\

\textbf{PR-AUC} &
$\displaystyle \int_{0}^{1} \text{Precision}(r)\,dr$ \\

\textbf{MCC} &
$\displaystyle
\frac{TP \cdot TN - FP \cdot FN}
{\sqrt{(TP + FP)(TP + FN)(TN + FP)(TN + FN)}}$ \\

\textbf{Log Loss} &
$\displaystyle -\frac{1}{N}\sum_{i=1}^{N}
\left[y_i \log(p_i) + (1 - y_i)\log(1 - p_i)\right]$ \\

\bottomrule
\end{tabular}
\end{table}

The target criteria were set to achieve accuracy above 80\% and a false positive rate below 20\%, reflecting practical real-world deployment thresholds.

\subsection{System Implementation Overview}

To support reproducibility and technical transparency, this section outlines the algorithmic design and key implementation components of the federated phishing-detection system. The implementation is modular, enabling efficient processing of large-scale web content while meeting privacy and performance requirements.
\begin{itemize}
    \item Overall System Workflow: The end-to-end pipeline comprises three phases: \emph{preparation}, \emph{training}, and \emph{deployment}.
    \item Memory-Efficient Data Processing: Chunk-based loaders handle large datasets without exhausting memory.
    \item Intelligent Vocabulary Construction: Filtering reduces the vocabulary from roughly 20 million tokens to 406{,}000, cutting noise while preserving coverage.
    \item CNN Training with Early Stopping: Training uses early stopping and continuous monitoring to stabilize and optimize performance.
    \item Federated Learning Framework: On-device updates enable privacy-preserving collaboration without sharing raw user data.
    \item Model Export and Mobile Deployment: The final model is optimized for mobile deployment while retaining federated-update capability.
\end{itemize}

The system is realized through a set of modular algorithms designed to ensure efficiency, accuracy, and seamless deployment within a federated learning paradigm. The overall workflow is coordinated by Algorithm~\ref{alg:main_workflow}, which organizes the process into three primary stages. Data preparation and preprocessing are handled in Algorithm~\ref{alg:prepare_data}, where raw inputs are cleaned, tokenized, and partitioned into federated subsets. The federated training loop, including client-side local updates and secure aggregation, is specified in Algorithm~\ref{alg:federated_training}. Finally, the conversion of the trained model for on-device execution and packaging for mobile environments is described in Algorithm~\ref{alg:deploy_model}. This structured separation improves clarity, reproducibility, and maintainability, while enabling straightforward future extensions such as alternative model architectures or privacy-enhancing mechanisms.

\begin{algorithm}[H]
\caption{Main Workflow of the Federated Phishing Detection System.}\label{alg:main_workflow}
\begin{algorithmic}
    \STATE \textbf{Input:} Raw website data $D_{raw}$; system configuration $P_{config}$
    \STATE \textbf{Output:} Deployed mobile model $M_{deployed}$

    \STATE \COMMENT{Phase 1: Data Preparation}
    \STATE $(\textit{Tokenizer}, D_{federated}) \leftarrow \textsc{PrepareData}(D_{raw})$ \hfill (\textbf{see Algorithm~\ref{alg:prepare_data}})

    \STATE \COMMENT{Phase 2: Federated Training}
    \STATE $M_{initial} \leftarrow \textsc{InitializeCNNModel}(P_{config})$
    \STATE $M_{trained} \leftarrow \textsc{FederatedTraining}(M_{initial}, D_{federated}, P_{config})$ \hfill (\textbf{see Algorithm~\ref{alg:federated_training}})

    \STATE \COMMENT{Phase 3: Deployment}
    \STATE $M_{deployed} \leftarrow \textsc{DeployModel}(M_{trained}, \textit{Tokenizer})$ \hfill (\textbf{see Algorithm~\ref{alg:deploy_model}})

    \STATE \textbf{return} $M_{deployed}$
\end{algorithmic}
\end{algorithm}

\begin{algorithm}[H]
\caption{Data Preparation.}
\label{alg:prepare_data}
\begin{algorithmic}[1]
    \STATE \textbf{Function} \textsc{PrepareData}($D_{raw}$)
    \STATE $D_{clean} \leftarrow$ clean, normalize, and balance $D_{raw}$
    \STATE \textit{Frequencies} $\leftarrow$ tokenize $D_{clean}$ and compute token frequencies
    \STATE \textit{Vocabulary} $\leftarrow$ filter \textit{Frequencies} using thresholds
    \STATE \textit{Tokenizer} $\leftarrow$ build tokenizer from \textit{Vocabulary}
    \STATE $D_{vectorized} \leftarrow$ convert $D_{clean}$ to sequences using \textit{Tokenizer}
    \STATE $D_{federated} \leftarrow$ partition $D_{vectorized}$ across client devices
    \STATE \textbf{return} (\textit{Tokenizer}, $D_{federated}$)
\end{algorithmic}
\end{algorithm}

\begin{algorithm}[H]
\caption{Federated Training Process.}
\label{alg:federated_training}
\begin{algorithmic}[1]
    \STATE \textbf{Function} \textsc{FederatedTraining}($M_{global}$, $D_{federated}$, $P_{config}$)
    \FOR{$r = 1$ to $P_{config}.\text{Rounds}$}
        \STATE $C_{selected} \leftarrow$ random subset of clients
        \STATE Updates $\leftarrow \emptyset$
        \FOR{each client $c \in C_{selected}$ \textbf{in parallel}}
            \STATE $M_{local} \leftarrow \textsc{TrainLocal}(M_{global}, D_c)$
            \STATE $\Delta_{local} \leftarrow M_{local} - M_{global}$
            \STATE append encrypted $\Delta_{local}$ to Updates
        \ENDFOR
        \STATE $\Delta_{agg} \leftarrow \textsc{Aggregate}(\textit{Updates})$ using FedAvg~\cite{mcmahan2017communication}
        \STATE $M_{global} \leftarrow M_{global} + \Delta_{agg}$
        \IF{\textsc{EarlyStoppingCriterionMet}()}
            \STATE \textbf{break}
        \ENDIF
    \ENDFOR
    \STATE \textbf{return} $M_{global}$
\end{algorithmic}
\end{algorithm}

\begin{algorithm}[H]
\caption{Model Deployment.}
\label{alg:deploy_model}
\begin{algorithmic}[1]
    \STATE \textbf{Function} \textsc{DeployModel}($M_{trained}$, \textit{Tokenizer})
    \STATE $M_{lite} \leftarrow$ convert $M_{trained}$ to TFLite (Float16 quantization)
    \STATE $T_{config} \leftarrow$ export tokenizer metadata
    \STATE $M_{package} \leftarrow \textsc{Package}(M_{lite}, T_{config})$ for mobile integration
    \STATE \textbf{Assert:} $M_{package}$ size $\leq 120$MB and latency $\leq 0.4$s
    \STATE \textbf{return} $M_{package}$
\end{algorithmic}
\end{algorithm}

\subsection{Federated Learning Implementation}

To ensure reproducibility and transparency, we report the key parameters and configuration details of the federated-learning setup.

\subsubsection{Simulation Environment}\label{subsec:flsim}
Experiments were conducted with TensorFlow Federated (TFF), an open-source framework by Google. The simulation consists of a single central server for aggregation and multiple client devices for local training, mirroring real-world deployment while remaining fully privacy-preserving.

\subsubsection{Federated Learning Process}

The federated workflow proceeds in five stages:
\begin{itemize}

    \item Global Model Distribution: The central server broadcasts the pretrained global weights to all participating clients.
    \item Local Model Training: Each client trains the model locally using only its own browsing history and the source code of visited pages. Weights are updated independently on-device from client-specific data.
    \item Secure Transmission of Updates: After local training, clients send only model updates (gradients or weight deltas) back to the server. No raw data or user-identifying information is shared, and all updates are encrypted in transit.
    \item Aggregation and Global Update: The server aggregates received updates via federated averaging (FedAvg) or a comparable method, producing a refreshed global model that reflects collective learning across clients \citep{onlu2025investigation}.
    \item Redistribution and Iteration: The updated global model is redistributed, and the cycle repeats. Iteration enables continual adaptation to new phishing tactics and evolving user behavior.
\end{itemize}

\subsubsection{Client Configuration and Data Distribution}

\begin{itemize}
    \item Number of Clients: The experiments used five simulated clients, emulating a small cohort of user devices under a controlled setup and matching the five Android devices used for the on-device evaluation (Section~\ref{subsec:flsim}).
    \item Data Partitioning: The training set was partitioned across clients under a Non-IID assumption to emulate real-world heterogeneity (e.g., differing phishing/legitimate ratios per client).
\end{itemize}

\subsubsection{Federated Training Protocol}

\begin{itemize}
    \item Aggregation Algorithm: FedAvg was used on the server side, computing a weighted average of client weights proportional to each client's local dataset size.
    \item Communication Rounds: Each round comprised global distribution, local client training, and server-side aggregation; training continued over successive rounds until an early-stopping criterion was met, and the federated results reported in Section~\ref{sec:results} correspond to the resulting aggregated global model over the five clients.
    \item Client Participation Rate: All 5 clients participated in every round (100\% participation).
\end{itemize}

\subsubsection{Local Client Training}

In each communication round, clients trained with the following configuration:
\begin{itemize}
    \item Local Epochs: In each round, every client performed four local training epochs on its own data---using the same optimizer and batch settings as the centralized model---to balance local adaptation with overfitting risk.
    \item Local Hyperparameters: Batch size of 32 and Adam optimizer with a learning rate of 0.0001, matching the centralized setup.
\end{itemize}

The federated learning paradigm offers two principal benefits: strong privacy guarantees---since no sensitive personal data ever leaves the user's device---and continuous adaptation with personalization. By training on a broad stream of up-to-date, real-world data, the model can learn device- and user-specific security patterns, improving robustness and accuracy against emerging phishing threats.

To further protect privacy and data integrity, all communication between clients and the server is encrypted via Transport Layer Security (TLS/SSL). In addition, cryptographic checks verify the integrity and authenticity of model updates throughout the pipeline. These technical and governance safeguards support compliance with applicable data-protection regulations, including the General Data Protection Regulation (GDPR) and Turkiye's Personal Data Protection Law (KVKK), thereby reinforcing user trust and system reliability.

\subsection{Integration into Anti-Phishing Browser}

The final CNN model was integrated into the open-source \emph{Lightning} web browser for Android \citep{lightning}. The anti-phishing module analyzes page source code in the background and issues real-time pop-up warnings when a threat is detected. The application presents a minimalist, user-friendly interface that clearly communicates security alerts.
\subsection{Performance Evaluation and Optimization}

\subsubsection{Performance Testing}

We evaluated speed, memory footprint, and accuracy across a range of Android devices. On average, per-page analysis completed in \(\approx 0.4\)~s, and the combined size of the model and tokenizer remained below \(120\)~MB (targets: \(<5\)~s per page, \(<200\)~MB storage). These results indicate that the deployment comfortably meets the real-time and storage constraints for mobile usage.

\subsubsection{Statistical Analysis and Optimization}

Collected metrics (e.g., accuracy, error rate) were compared across device cohorts to assess robustness under heterogeneous hardware conditions. Based on these findings and user feedback, we iteratively refined hyperparameters, network architecture, and software components, yielding consistent accuracy gains without compromising latency or on-device resource usage.

\subsection{Security and Privacy Measures}

\begin{itemize}
    \item Encrypted Communication: All client--server exchanges employed TLS/SSL to protect confidentiality and integrity.
    \item Legal Compliance: The system design aligns with GDPR, KVKK, and other applicable regulations, supporting responsible data handling and user privacy.
\end{itemize}

\section{Experimental Results and Discussion}\label{sec:results}

This section reports a comprehensive evaluation of the proposed deep learning--based phishing detector. We analyze performance using accuracy, precision, recall, F1-score, ROC-AUC, and complementary metrics, and we summarize the training dynamics, data characteristics, and statistical findings with supporting tables and figures.

\subsection{Properties of Training and Test Datasets}

The training corpus comprises 100{,}000 phishing and 49{,}524 legitimate website source codes. After class balancing, we used 34{,}666 samples per class for training. The test set contains 8{,}999 phishing and 8{,}999 legitimate samples (\(17{,}998\) total). Drawing from multiple independent sources increases robustness to real-world variability in site distributions and content.

\begin{table*}
\caption{Sources and Sizes of Training and Test Datasets}
\label{tab:datasources}
\centering
\resizebox{\linewidth}{!}{
\begin{tabular}{llll}
\toprule
\textbf{Type} & \textbf{Source} & \textbf{Train (raw/sample)} & \textbf{Test (raw/sample)} \\
\midrule
Phishing  & PhishTank, OpenPhish, PhishRepo  & 100{,}000 (34{,}666) & 29{,}997 (8{,}999) \\
Legitimate & Alexa Top Sites, Webpage Rank, Google & 49{,}524 (34{,}666) & 49{,}990 (8{,}999) \\
\bottomrule
\end{tabular}
}
\end{table*}

\subsection{Word Distribution and Tokenization Analysis}

After preprocessing and filtering, only 406{,}060 of the 19{,}971{,}995 unique tokens were retained. This reduction shrinks the tokenizer's vocabulary and helps the model emphasize informative features over noise. The most frequent tokens predominantly correspond to HTML, Cascading Style Sheets (CSS), and JavaScript structures, suggesting that the model captures structural distinctions between legitimate and phishing pages.

\begin{table}
\centering
  \caption{Word distribution and filtering results.}
  \label{tab:worddist}
  \begin{tabular}{lc}
    \toprule
    Metric & Count \\ \midrule
    Total unique words & 19,971,995 \\
    Words used after filtering & 406,060 \\
    Filtered-out word ratio & 97.97\% \\ \bottomrule
  \end{tabular}

  {\footnotesize Top-10 frequent tokens: \texttt{class}, \texttt{div}, \texttt{span}, \texttt{href}, \texttt{https}, \texttt{data}, \texttt{com}, \texttt{quot}, \texttt{width}, \texttt{content}.}
\end{table}

\subsection{Model Training and Learning Curves}

\begin{figure}[ht]
  \centering
  \includegraphics[width=0.80\linewidth]{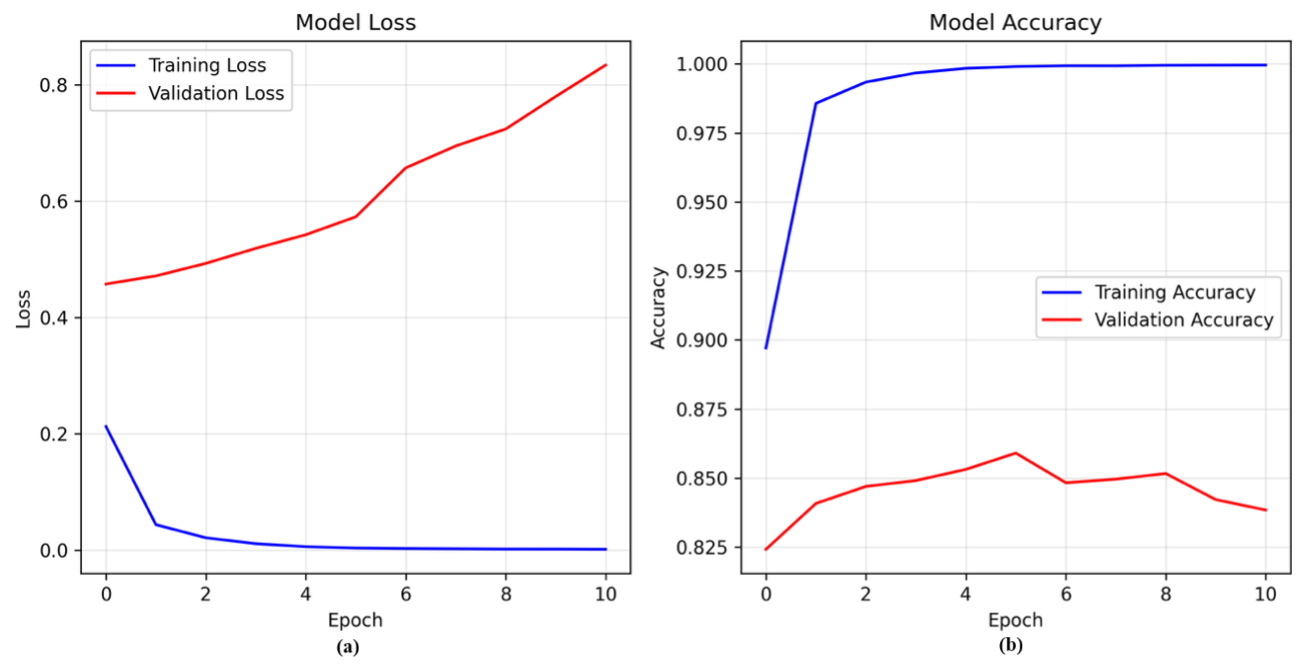}
  \caption{Training and validation loss/accuracy curves.}
  \label{fig:trainingcurves}
\end{figure}

Figure~\ref{fig:trainingcurves} shows the evolution of training and validation metrics across epochs. The model rapidly reaches high training accuracy with low loss. A subsequent rise in validation loss indicates some overfitting---expected given distributional differences between training and test data---yet the overall trajectory demonstrates good generalization to unseen, real-world samples.

\subsection{Classification Performance and Confusion Matrix}

\begin{figure}[ht]
\centering
\includegraphics[width=0.50\linewidth]{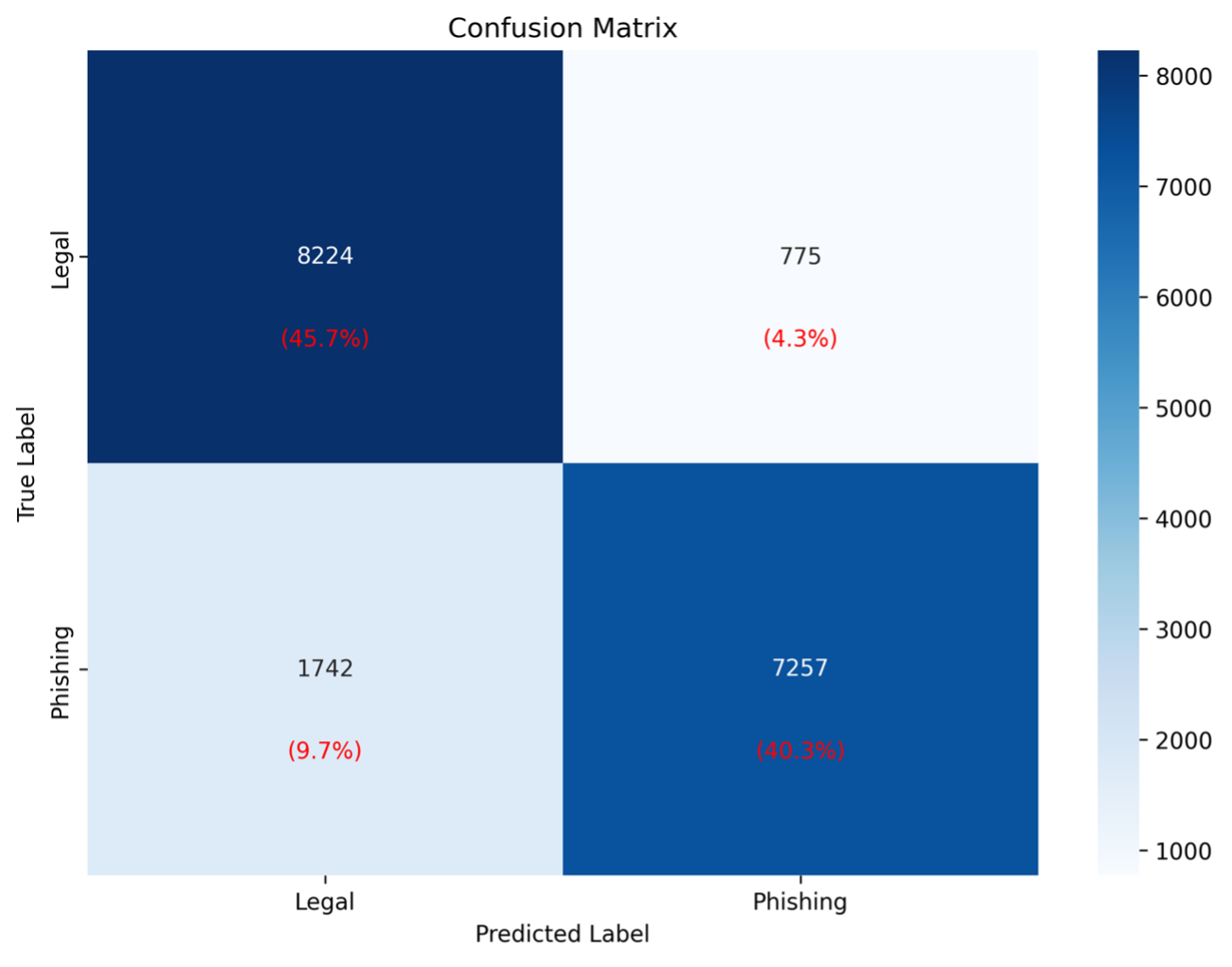}
\caption{Confusion matrix of the test set.}
\label{fig:confmatrix}
\end{figure}

Table~\ref{tab:testmetrics} reports test-set performance, and Figure~\ref{fig:confmatrix} visualizes the confusion matrix. The model attains \textbf{86.02\%} accuracy, \textbf{90.35\%} precision, \textbf{80.64\%} recall, \textbf{85.22\%} F1-score, and \textbf{92.78\%} ROC-AUC, indicating reliable separation of phishing and legitimate pages with a low false-positive rate.

\begin{table}
\centering
  \caption{Performance metrics on the test set.}
  \label{tab:testmetrics}
  \begin{tabular}{lc}
    \toprule
    Metric & Value \\ \midrule
    Accuracy & 0.8602 \\
    Precision & 0.9035 \\
    Recall & 0.8064 \\
    F1-score & 0.8522 \\
    Specificity & 0.9139 \\
    ROC-AUC & 0.9278 \\
    PR-AUC & 0.9230 \\
    Matthews correlation & 0.7245 \\
    Log loss & 0.5730 \\ \bottomrule
  \end{tabular}
\end{table}

From Figure~\ref{fig:confmatrix}, the classifier correctly identifies \(8{,}224\) legitimate sites and mislabels \(775\) as phishing; for phishing pages, it detects \(7{,}257\) correctly and misses \(1{,}742\) (i.e., classifies them as legitimate). These counts correspond to a false-positive rate of \(1 - \text{Specificity} = 0.0861\) and a false-negative rate of \(1 - \text{Recall} = 0.1936\), supporting the model's practical suitability for real-world mobile security applications.

\subsection{ROC and Precision--Recall Curves}

\begin{figure}[H]
\centering

\begin{subfigure}{0.48\linewidth}
    \centering
    \includegraphics[width=\linewidth]{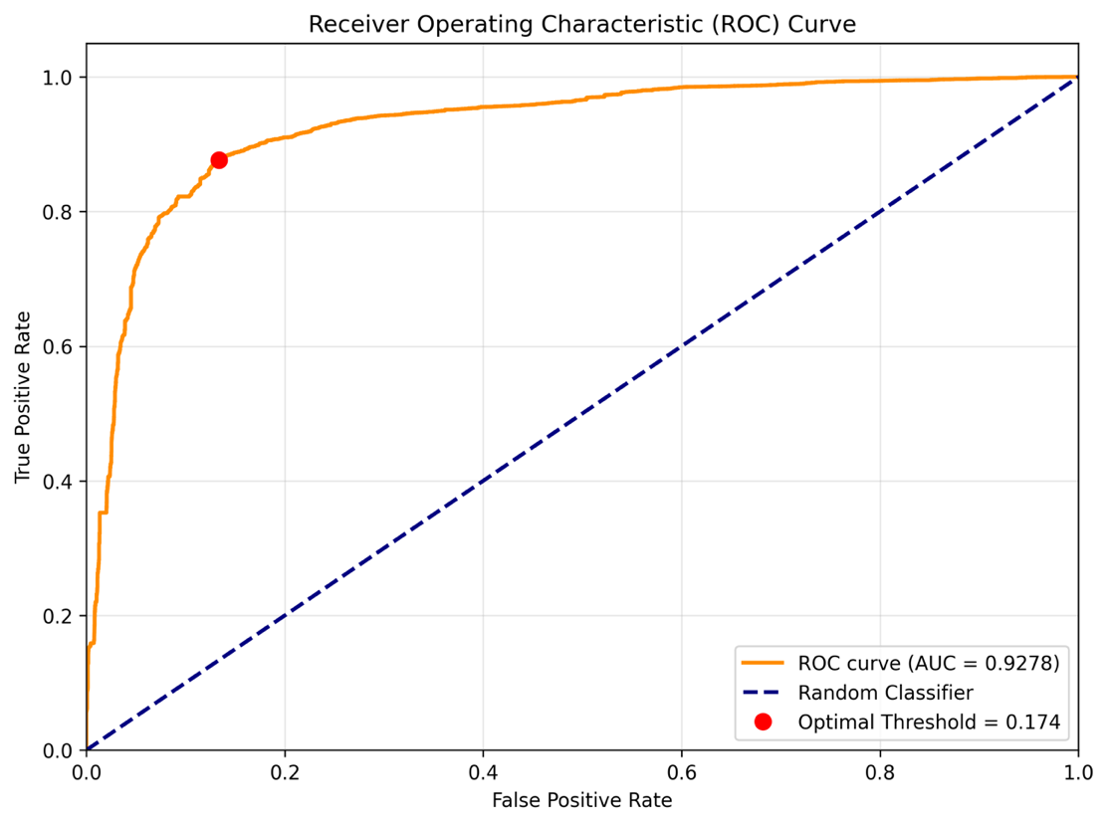}
    \caption{ROC curve (\(\text{AUC}=0.9278\))}
    \label{fig:roc}
\end{subfigure}
\hfill
\begin{subfigure}{0.48\linewidth}
    \centering
    \includegraphics[width=\linewidth]{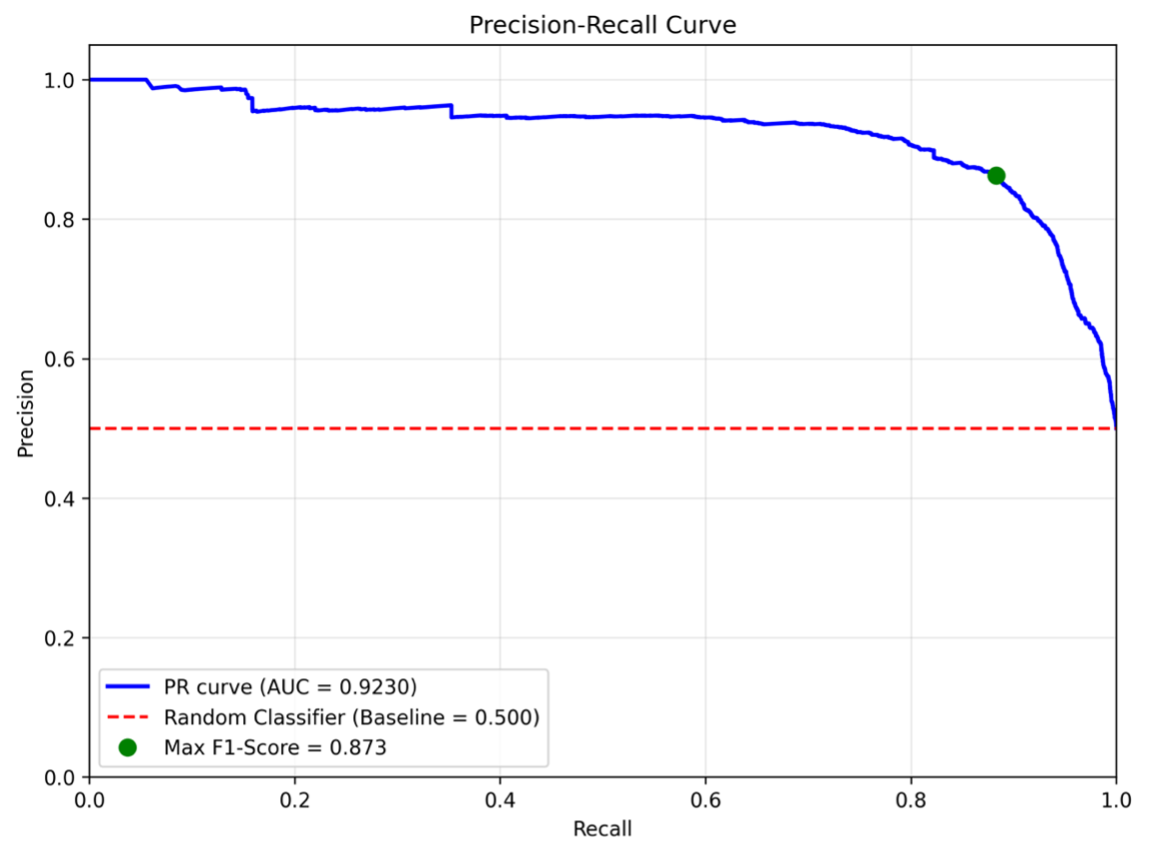}
    \caption{PR curve (\(\text{AUC}=0.9230\))}
    \label{fig:pr}
\end{subfigure}

\caption{Performance curves: (a) ROC and (b) Precision--Recall.}
\label{fig:roc_pr}
\end{figure}

The ROC and Precision--Recall (PR) curves in Figures~\ref{fig:roc} and \ref{fig:pr} illustrate the model's discriminative ability across thresholds. The areas under the curves---\(\text{ROC-AUC}=0.9278\) and \(\text{PR-AUC}=0.9230\)---indicate strong separability and reliable performance, including under class-imbalance conditions.

\subsection{Prediction Analysis and Model Calibration}

\begin{figure}[H]
  \centering
  \includegraphics[width=0.95\linewidth]{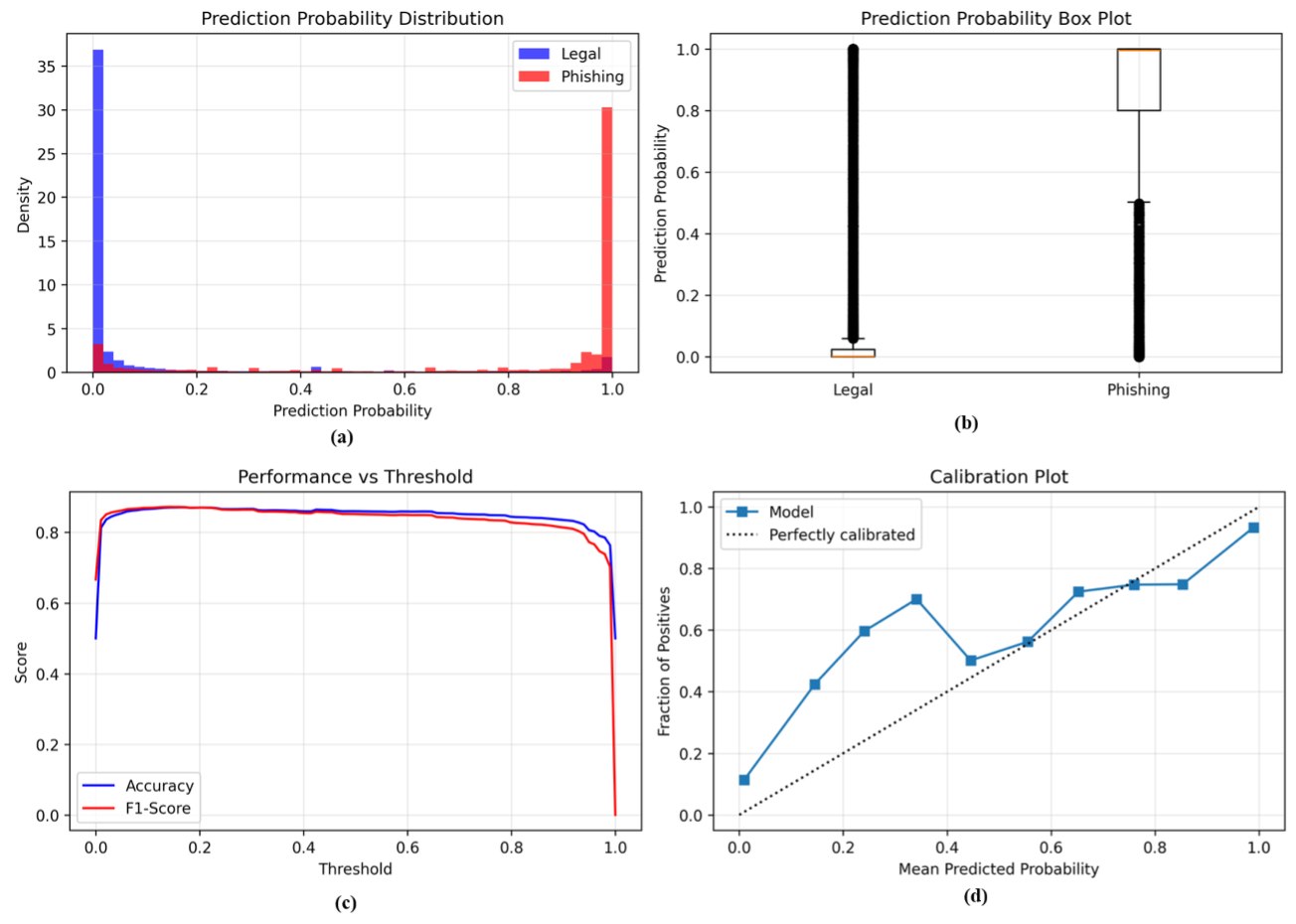}
  \caption{Prediction probability distributions, threshold analysis, and calibration results.}
  \label{fig:calibration}
\end{figure}

Figure~\ref{fig:calibration} summarizes prediction score distributions, box plots, performance-versus-threshold trends, and a calibration curve. The model generally assigns high probabilities to the correct class, and the calibration plot suggests that predicted probabilities are well aligned with observed frequencies, albeit not perfectly calibrated.

\subsection{Model Size and Device Performance}

The final model occupies \(\,99.38\,\text{MB}\) (with tokenizer: \(\,116.52\,\text{MB}\)). Tests across five Android devices show an average per-page inference time of \(\approx 0.4\)~s, confirming suitability for real-time mobile use.

\subsection{Federated Learning Performance and Comparison with the Initial Model}

To assess the benefits of federated learning, updates from five client devices were aggregated on the central server to produce a new global model. We then compared this federated model to the initial model---the centrally trained, Bayesian-optimized CNN without federation---over key metrics (Table~\ref{tab:federatedvsinitial}). For reference, the table also reports an unoptimized CNN baseline (without Bayesian hyperparameter optimization), which performed roughly 3--5\% below the optimized model across metrics, quantifying the contribution of the optimization step. Overall, federated training improves accuracy, recall, F1, ROC-AUC, and PR-AUC, with a small trade-off in precision---consistent with a model that is slightly more sensitive to phishing while maintaining a low false-positive rate.

\begin{table}
\centering
  \caption{Configuration ablation across the pipeline: an unoptimized CNN, the centrally trained Bayesian-optimized CNN (the initial model), and the federated global model (after FedAvg aggregation). The last column reports the relative change of the federated model over the optimized CNN.}
  \label{tab:federatedvsinitial}
  \begin{tabular}{lcccc}
    \toprule
    Metric & Unoptimized CNN & Optimized CNN & Federated CNN & Rel.\ change (\%) \\ \midrule
    Accuracy  & 0.8249 & 0.8602 & 0.8776 & +2.0 \\
    Precision & 0.8719 & 0.9035 & 0.8960 & $-0.8$ \\
    Recall    & 0.7701 & 0.8064 & 0.8251 & +2.3 \\
    F1-score  & 0.8178 & 0.8522 & 0.8589 & +0.8 \\
    ROC-AUC   & 0.8926 & 0.9278 & 0.9431 & +1.6 \\
    PR-AUC    & 0.8850 & 0.9230 & 0.9385 & +1.7 \\ \bottomrule
  \end{tabular}
\end{table}

As summarized in Table~\ref{tab:federatedvsinitial}, the federated model yields clear gains over the initial baseline---most notably in accuracy (\textbf{+2.0\%}), recall (\textbf{+2.3\%}), and ROC-AUC (\textbf{+1.6\%}). These improvements indicate that aggregating updates from heterogeneous clients enhances generalization and strengthens robustness to real-world phishing attempts. The higher recall, in particular, reflects more effective detection of phishing sites, which is critical for end-user protection.

A modest drop in precision (\textbf{--0.8\%}) suggests a slight increase in false positives; however, this trade-off is outweighed by the broader performance gains and does not materially undermine practical reliability. Increases in F1-score and PR-AUC further confirm a better balance between sensitivity and specificity.

\subsection{Comparison with Literature and Applicability}

To the best of our knowledge, no prior work performs \emph{real-time} phishing detection on mobile devices by jointly leveraging full web page source code, CNN-based feature extraction, and federated learning (FL). Existing mobile solutions typically rely on shallow features (e.g., URLs, HTML tags, form attributes) and classical machine learning models (e.g., Decision Trees, SVMs, Random Forests), often reporting accuracies above 90\%. However, these approaches neither exploit deep representations of full source code nor address privacy and personalization through FL.

In contrast, the proposed system enables (i) code-centric, automatic feature extraction via CNNs and (ii) privacy-preserving, adaptive learning through FL in a real-time mobile setting. Although the achieved accuracy (86.02\%) is slightly lower than some classical baselines, this reflects the increased complexity of full-source-code analysis and the constraints of on-device deployment. Importantly, FL enhances privacy while enabling continual adaptation to evolving attacks and heterogeneous user behavior.

Prior deep learning studies are typically limited to centralized, desktop environments, whereas mobile implementations remain shallow-feature-based. This work bridges that gap by integrating deep, code-level analysis with federated, real-time operation on mobile platforms.

Overall, the proposed framework advances the literature by combining deep learning and FL into a practical, privacy-aware, and adaptive mobile phishing detection system.

\section{Conclusion and Discussion}\label{sec:conclusion}

We present a real-time, privacy-preserving phishing detection system that analyzes full web page source code using a CNN and is continuously updated via FL~\citep{demir2026decentralized}. The approach enables on-device training without transmitting user data, ensuring strong privacy while maintaining generalization across heterogeneous data sources.

The model achieved \(86.02\%\) accuracy, \(90.35\%\) precision, and \(92.78\%\) ROC-AUC on large-scale datasets. Deployment within the open-source \emph{Lightning} browser on Android demonstrated practical feasibility, with an average inference latency of \(0.4\)~s and a total package size below \(120\)~MB. While the accuracy is slightly lower than some centralized baselines, the system prioritizes a balanced trade-off between accuracy, privacy, real-time performance, and deployability. Notably, accuracy improved to \(87.76\%\) after one FL round, highlighting its adaptability to evolving threats.

The main contribution is the integration of (i) code-level feature extraction via CNNs, (ii) real-time mobile inference, (iii) privacy preservation through FL, and (iv) continuous model updates---capabilities typically addressed separately in prior work.

Although evaluated primarily on mobile devices, the lightweight and decentralized design suggests applicability to resource-constrained environments such as IoT. The FL framework further aligns with data protection regulations (e.g., GDPR, KVKK), making it suitable for privacy-sensitive deployments.

Limitations include evaluation in a simulated FL setting. Future work should validate the system at scale in real-world environments, incorporate advanced privacy mechanisms (e.g., differential privacy), and extend platform support. Overall, the proposed system demonstrates that real-time, AI-driven phishing detection can provide an effective and privacy-aware alternative to static blacklist-based approaches.

\section{Future Work}\label{sec:future}

Future research will focus on improving model capacity, scalability, and deployment flexibility:

\begin{itemize}
\item \textbf{Advanced Architectures:} Explore Transformers and Graph Neural Networks (GNNs) while maintaining low latency and memory usage.
\item \textbf{Cross-lingual Generalization:} Extend evaluation across languages and regions.
\item \textbf{Attack Diversity:} Improve detection of varied phishing types (e.g., spear-phishing, smishing).
\item \textbf{Platform Expansion:} Support iOS, IoT, and embedded systems.
\item \textbf{Continual Learning:} Incorporate real-world feedback to reduce false positives/negatives.
\item \textbf{Error Analysis:} Analyze misclassifications (e.g., obfuscation, adversarial inputs).
\item \textbf{Security Enhancements:} Strengthen FL against poisoning and inference attacks using secure aggregation and differential privacy.
\item \textbf{Open Deployment:} Promote adoption via open-source and public-sector collaborations.
\end{itemize}

\section*{Disclosure statement}
The authors declare that they have no conflict of interest.

\section*{Author Contribution}
Eren Dogan; Conceptualization, Data curation, Formal analysis, Investigation, Methodology, Resources, Software, Writing - original draft, Writing - review and editing.
Busra Buyuktanir; Supervision, Writing - original draft, Writing - review and editing.
Dr. Gozde Karatas-Baydogmus; Supervision, Writing - original draft, Writing - review and editing.
Dr. Kazim Yildiz; Supervision, Writing - original draft, Writing - review and editing.

\section*{Acknowledgments}
This work was supported by TUB\.{I}TAK under the 2209-A Research Project Program.

\bibliographystyle{plainnat}
\bibliography{references}

\end{document}